\documentclass[twocolumn]{aastex631}

\newcommand{\kms}{\rm km~s^{-1}}
\newcommand{\cmc}{\rm cm^{-3}}
\newcommand{\zsun}{\rm Z_\odot}
\usepackage{natbib}
\usepackage{amsmath}
\usepackage{CJK}
\usepackage{multirow}
\usepackage{booktabs}
\usepackage{makecell}
\usepackage{todonotes}
\usepackage{savesym}
\savesymbol{tablenum}
\usepackage{siunitx}
\restoresymbol{SIX}{tablenum}

\newcommand{\rion}[2]{{\ensuremath{\mbox{\rm #1$\,${\small\uppercase\expandafter{\romannumeral#2\relax}}}}}}
\newcommand{\HII}{\rion{H}{2}}
\newcommand{\Ha}{\ensuremath{\mbox{\rm H$\alpha$}}}
\newcommand{\Hb}{\ensuremath{\mbox{\rm H$\beta$}}}

\newcommand{\OIII}{[\rion{O}{3}]}
\newcommand{\OII}{[\rion{O}{2}]}
\newcommand{\OI}{[\rion{O}{1}]}
\newcommand{\NII}{[\rion{N}{2}]}
\newcommand{\SII}{[\rion{S}{2}]}
\newcommand{\SIII}{[\rion{S}{3}]}
\newcommand{\FeII}{[\rion{Fe}{2}]}

\newcommand{\OIHa}{\OI/\Ha}

\newcommand{\NIIHa}{\NII/\Ha}
\newcommand{\SIIHa}{\SII/\Ha}

\newcommand{\OIIIHb}{\OIII/\Hb}

\newcommand{\M} {\ensuremath{\mbox{\rm \texttt{MAPPINGS}}}}

\newcommand{\MIII}{\ensuremath{\mbox{\rm \texttt{MAPPINGS\,III}}}}

\newcommand{\MV}{\ensuremath{\mbox{\rm \texttt{MAPPINGS\,V}}}}

\begin{document}
\begin{CJK*}{UTF8}{gbsn}

\title{ Effects of Preionisation in Radiative Shocks IV. A Spectral Grid Library of fully Radiative Shock Models with Time--Dependent Precursor\footnote{This paper is in memory of Mike A. Dopita}\footnote{The full library is available in Zenodo at
\url{https://doi.org/10.5281/zenodo.20716483}}}

\author[0000-0003-0401-3688]{Yifei Jin（金刈非）}
\email{yfjsci@gmail.com}
\affiliation{Westlake University, 
600 Dunyu Road, Xihu District, 
Hangzhou, Zhejiang 310030 PR China}

\author[0000-0002-4134-864X]{Knox S. Long}
\affiliation{Space Telescope Science  Institute, 
3700 San Martin Drive, 
Baltimore, MD 21218 USA}

\author[0000-0002-6620-7421]{Ralph Sutherland}
\affiliation{Australian National University,
Cotter Road, 
Canberra, ACT 2611 Australia}

\author[0000-0002-8614-6275]{Yong Shi（施勇）}
\affiliation{Westlake University, 
600 Dunyu Road, Xihu District, 
Hangzhou, Zhejiang 310030 PR China}

\begin{abstract}

Here we present an extensive set of shock models calculated using an improved version of the \M\ shock and photoionization code. The new version of \M, namely \MV, has a greatly improved treatment of the precursor emission, a more robust adaptive-mesh shock solver \cite{Sutherland2017}, and new atomic data compared to \MIII, which was used by  \cite{Allen2008} to create their widely used shock model grid.  The new shock model library covers the shock velocity range $50-1000\;\kms$, and a wide range of initial conditions and updated abundances.  We highlight how the newer models compare to the earlier grid, and discuss where the shock models are most likely to approximate conditions for various types of astrophysical objects.  The emission that one expects to observe depends on a variety of factors, including the velocity of the shock, the density and abundance of material the shock is encountering, the magnetic field and the effective age of the shock.

\end{abstract}


\section{Introduction} \label{sec:intro}

Shocks arise when material and/or magnetic fields are injected into the ISM at supersonic speeds.  
Material passing through the shock interface is heated and compressed behind the shock, and emits radiation over a large portion of the electromagnetic spectrum as the material is ionized, and ultimately recombines \citep{Shull1979}.  
Additionally, radiation can arise from a shock precursor, material which has not yet crossed the shock front, but which is photoionized by X-rays and EUV emission arising in the post shock gas \citep{Dopita2017,Raymond2026}.
The observed spectrum depends on a variety of factors, including the shock velocity, the density and abundances of the material the shock is encountering, the ambient magnetic field, and the effective age of the shock \citep{Sankrit2014,Dopita2019}. 

Here we are focusing our efforts on so-called complete shocks. 
They are complete in that we will be calculating the integrated radiation spectrum for a plasma that has been heated and then allowed to cool back to the temperature of the interstellar medium.  
Such shocks are also termed radiative, because radiative cooling strongly affects the structure of shock \citep{Cox1972,Raymond1976,Dopita1976}.
These are the shocks that are most often seen at optical wavelengths, where the bulk of the radiation arises from the Balmer lines of H, or forbidden lines of abundant ions such as \NII, \OI, \OII, \OIII, {[\rion{S}{1}]}, \SII, \SIII\ and \FeII.

There is a long history of attempts to model shocks and to compare shock models to optical observations \citep{Cox1972,Dopita1976,Raymond1976,Shull1979, Binette1985}, UV observations \citep{Hartmann1984}, and X-ray observations \citep{Borkowski2001,Reynolds2024}.
All of these modelling efforts assume a 1D plane parallel geometry, and for the most part steady state conditions, although a few considered situations where the plasma did not fully recombine (so-called incomplete shocks) and a few others considered the weak permitted-line only emission that is observed in shocks that are ionizing (radiationless shocks). As time has passed, the sophistication of the modelling has increased, with better atomic data and more physical processes considered.

The usage of shock models are twofold. 
The first way is to constrain the physical parameters of the interstellar medium in SNRs.
\cite{Dopita2016,Dopita2018,Dopita2019} used shock models to compare to the optical emission-line spectra to calibrate the shock velocity, preshock gas density, the chemical abundance of ISM and the dust composition of the SNRs in Large Magellanic Cloud.
\cite{Raymond2020a,Raymond2020b,Raymond2023,Raymond2025} used models to delicately measure the ram pressure, magnetic field strength and the turbulent vorticity of shock waves in the Cygnus Loop. 
Conversely, grids of shock emission-line ratios are used in combination with the ratios of photoionization models, to separate the dominant ionization mechanisms in external galaxies between photoionization of stars or shock waves of SNRs \citep{Sanchez2014,Juneau2022,Zheng2024}.

In recent years, most of the comparisons to large samples have been made by using the \cite{Allen2008} shock grids (hereafter \MIII\ grids), which were created using the \MIII\ code. 
In \MIII\  and hence in both of these grids, the precursor and post-shock region are computed separately and the precursor is treated as a photoionization equilibrium gas.
This treatment assumes that the precursor gas is exposed to a mean upstream EUV field and the cooling happens immediately after the ionization. 
More recently, \cite{Alarie2019} used an early version of \MV~(\texttt{v5.1.13}) to construct a set of grids but this version didn't include a proper treatment of precursors as is included in the latest \MV~(\texttt{v5.2.1}).

For the latest \MV, \cite{Sutherland2017} implemented a multi-zone approach to solve the precursor structures.
This allows for a fully self-consistent calculation of the shock-precursor flow covering the structures before and after the shock front.
Here, we present a new shock library with time-dependent shock-precursor spectra by using $\MV$.
We describe the $\MV$ and the shock models in Section~\ref{sec:mapv} and \ref{sec:model}.
We describe the shock grids in Section~\ref{sec:grid}.
The structure of the post-shock and precursor is described in Section~\ref{sec:postshock} and \ref{sec:precursor}.
The shock and precursor emission-lines are presented in Section~\ref{sec:line}.
We present the comparison between the \MIII\ and \MV\ grids in Section~\ref{sec:comp_MIII}, and the comparison between the \MV\ grids and observations in Section~\ref{sec:comp_sdss} and \ref{sec:comp_snr}.
Finally, we summarize our results and conclusions in Section~\ref{sec:conclusion}.

\section{Mappings\,V Description}\label{sec:mapv}

\MV\ \citep{Sutherland2017} is the latest version of the \M\ code, which includes the CHIANTI~10 database of atomic data, providing improved modeling of thermally ionized and photoionized plasmas, especially the X-ray spectra of nonequilibrium plasmas. 
The number of cooling and recombination lines is expanded from $\sim$2000 in previous versions to 80,000 in the current code.

\section{How shock models are calculated} \label{sec:model}

The \MV\ code\footnote{The updated version of \MV\ can be found at https://github.com/MAPPINGS-Code/MAPPINGS-V} is designed to calculate both shock and photoionization processes. 
The shocks are described by the Rankine-Hugoniot flow equations and the pressure is the principal flow variable as it is the driving force in all shock phenomena.
Following \cite{Cox1972}, both the steady shocked flow and the shock jump are solved in the same manner by deriving the solution to the compression factor, $x\equiv\rho_1/\rho_0$, between any two points in the flow.
The entire shocked flow can be consistently computed by using a single shock flow solver.

First, the shocked flows maintain the conservation of mass:
\begin{equation}
    \rho_0 v_0 = \rho_1 v_1,
\end{equation}
where $\rho$ is the gas density and $v$ is the gas velocity. 
The subscript number refer to any two points, $f_0$ and $f_1$, in the flow.
The compression factor, $x$, is defined as
\begin{equation}
    x \equiv \rho_1/\rho_0 = v_0/v_1.
\end{equation}
The transverse magnetic field perpendicular to the flow velocity is constrained by:
\begin{equation}
    B_0/\rho_0 = B_1/\rho_1.
\end{equation}
There are three pressures, gas pressure $P_{gas}$, ram pressure $P_{ram}$, and magnetic pressure $P_B$ in shocks:
\begin{equation}
    P_gas = nkT = \frac{\rho kT}{\mu m} \equiv (\gamma -1)E,
\end{equation}
\begin{equation}
    P_{ram} = \rho v^{2},
\end{equation}
\begin{equation}
    P_B = B^2 / 8\pi,
\end{equation}
where $k$ is the Boltzmann constant, $T$ is the temperature, $n$ is the total gas number density, $\mu$ is the molecular weight per gas particle, $m$ is the atomic mass unit, and $E$ is the internal specific energy.
The conservation of momentum flux is expressed in terms of the pressures:
\begin{equation}
    P_{gas,0} + P_{ram,0} + P_{B,0} = P_{gas,1} + P_{ram,1} + P_{B,1}.
\end{equation}
Finally, the energy conservation is expressed as:
\begin{equation}
\begin{split}
        \frac{1}{\rho_0}\left(gP_{gas,0} + \frac{1}{2}P_{ram,0} + 2P_{B,0} - \langle L \rangle \Delta t \right) \\
= \frac{1}{\rho_1}\left(gP_{gas,1} + \frac{1}{2}P_{ram,1} + 2P_{B,1} \right),    
\end{split}  
\end{equation}
where $g=\gamma/(\gamma-1)$, and $\langle L \rangle \Delta t$ is the energy loss between $f_0$ and $f_1$ through the energy radiated from shocks.

\subsection{The precursor calculation}
For shocks with velocity $v_{shock}>40 \rm ~km~s^{-1}$ \citep{Dopita2017}, the upstream EUV ionizing field forms in the precursor medium, determined by the escaping UV photons produced in the cooling zone in postshock regions.
The photoionization in the precursor changes from equilibrium to non-equilibrium ionizations as the shock velocity declining from fast shocks ($v_{shock}>$100~km~s$^{-1}$) to slow shocks ($v_{shock}\sim$50~km~s$^{-1}$).
In previous fast shock grids \citep{Allen2008}, precursors are assumed to be in pure equilibrium with constant density and computed separately from the shocked flow.
In this paper, the precursor is computed consistently at all velocities by solving the time-dependent equilibrium and nonequilibrium photoionizations, and the radiative transfer of ionizing photons back-and-forth over the shock front.

\section{Model Grid Input Parameters}\label{sec:grid}

Shock models are determined by the preshock density $n_H$, the shock velocity $v_{shock}$, the preshock magnetic field that is transverse to shock fronts $B$, and the chemical abundance of ISM.
We have calculated two new sets of \MV\ models: a set that we feel should be used by researchers for comparing with observations going forward, which covers a larger set of shock conditions than \MIII\, and which reflects the latest chemical abundances, and an \MIII\ set which replicates exactly the parameters used by  \cite{Allen2008} in order to permit direct \MIII\ - \MV\ comparisons.   %

\begin{table*}
\centering
\caption{Elemental abundances (log$_{10}$ [X/H]) used in the MAPPINGS III fast shock models of \cite{Allen2008}.}
\begin{tabular}{lcccc}
\hline
Element & Solar & 2$\times$Solar & LMC & SMC \\
\hline
H  & 0.00  & 0.00  & 0.00  & 0.00  \\
He & -1.01 & -1.01 & -1.05 & -1.09 \\
C  & -3.44 & -3.14 & -3.96 & -4.24 \\
N  & -3.95 & -3.65 & -4.86 & -5.37 \\
O  & -3.07 & -2.77 & -3.65 & -3.97 \\
Ne & -3.91 & -3.61 & -4.39 & -4.73 \\
Na & ---   & ---   & -4.85 & -5.92   \\
Mg & -4.42 & -4.12 & -4.53 & -5.01 \\
Al & -5.53 & -5.23 & -4.28 & -5.60 \\
Si & -4.45 & -4.15 & -5.29 & -4.69 \\
S  & -4.79 & -4.49 & -7.23 & -5.41 \\
Cl & ---   & ---   & ---   & -7.30  \\
Ar & -5.44 & -5.14 & -5.71 & -6.29 \\
Ca & -5.88 & -5.58 & -6.03 & -6.16 \\
Fe & -4.63 & -4.33 & -4.77 & -5.11 \\
Ni & ---   & ---   & -6.04 & -6.14   \\
\hline
\end{tabular}
\end{table*}

\subsection{The MAPPINGS\;V Shock Model Library}

As indicated in Table~\ref{tab:MV_parameter}, the \MV\ shock model library covers the magnetic field $B$ ranges from $10^{-4}$ to $100~\mu$G, densities from 0.01 to 100 cm$^{-3}$, and shock velocities $v_{\rm s}$  from 50 to 1000 km s$^{-1}$ in steps of 25 km s$^{-1}$. The abundances we use are from the Galactic Concordance abundance set \citep{Nicholls2017} as listed in Table~\ref{tab:abundance}.

\begin{table*}
\centering
\caption{Parameter space of the MAPPINGS V shock model grid.}
\label{tab:shock_grid}
\begin{tabular}{ll}
\hline
Parameter & Values \\
\hline
Magnetic field, $B$ ($\mu$G) & 0.0001, 0.001, 0.01, 0.1, 0.5, 1.0, 2.0, 3.23, 4.0, 5.0, 8.0, 10.0, 50.0, 100.0 \\
preshock density, $n_{H}$ (cm$^{-3}$) & 0.01, 0.1, 1.0, 10, 100 \\
Shock velocity, $v_{\rm s}$ (km s$^{-1}$) & 50--1000, in steps of 25 \\
Chemical abundance, Z (Z$_\odot$) & 0.05, 0.2, 0.4, 1.0, 2.0 (Galactic Concordance; \citealt{Nicholls2017}) \\
\hline
\end{tabular}\label{tab:MV_parameter}
\end{table*}

\subsubsection{Magnetic field strength}

Magnetic field strength varies broadly in different circumstances of interstellar medium.
In our Milky Way galaxy with moderate density and pressures in the disk of $\log P/k \sim 4$, we observe a range of magnetic fields, largely covered by the \MIII\ grid.  
If pressures result from interactions and equipartition processes, then larger and smaller or different classes of galaxies may have a wider range of magnetic fields, so we take the observed Milky Way magnetic fields and extend the grid to lower and higher magnetic fields to cover other objects, and reduce the need for extrapolations in Milky Way like galaxies but with different magnetic field ranges.  

In the Milky Way, the magnetic field strengths are observed to be $B = 8.0$ - 100~$\mu$G in supernova remnants \citep{Tran2015, Han2017}, and 2--6~$\mu$G in the Galactic H\,\textsc{ii} regions \citep{Harvey-Smith2011, Purcell2015, Savage2013, Planck2016}.
We start with the $B=1-10\mu$G, similar to the observations of the Milky Way, and add the lower and higher magnitude values to cover a wider range of objects for the convenience of exploring the ISM with the unknown extreme magnetic field strengths. 
In the current library, the lowest and highest magnetic field strengths reach to 10$^{-4}~\mu$G and 100~$\mu$G as listed in Table~\ref{tab:MV_parameter}.

\subsubsection{Preshock density}

The preshock density also spans a wide range across different astrophysical environments. 
In the Cygnus Loop, the preshock densities vary from 1 to 100~cm$^{-3}$ \citep{Raymond2020b}. 
The $n_{\rm H} = 1.0$, 10, and 100~cm$^{-3}$ models cover this regime of supernova remnant shocks. 
For shocks in the narrow-line regions around AGNs, the preshock density is of order a few hundred cm$^{-3}$ \citep{Dopita2015, Davies2020}, which is represented by the $n_{\rm H} = 100~\mathrm{cm}^{-3}$ models. 
In the local ISM, the density can be as low as 0.1~cm$^{-3}$, with an electron density of 0.01~cm$^{-3}$ \citep{Slavin2008, Linsky2019}.
The $n_{\rm H} = 0.01$ and $0.1~\mathrm{cm}^{-3}$ models are applicable to this diffuse, low-density regime.

\subsubsection{Shock velocity}

Shock emissions are observed in galactic outflows, supernova remnants, and Herbig-Haro objects, spanning a wide range of velocities. 
In outflows, typical shock velocities are of order hundreds of km~s$^{-1}$ \citep{Heckman1990, Veilleux2005, Ubler2023}, reaching speeds of up to 1000~km~s$^{-1}$ in intense starburst galaxies and quasars \citep{Strickland2009, Cresci2023}. In nearby optical supernova remnants (SNRs), measured shock velocities are $\sim$200~km~s$^{-1}$ in the Cygnus Loop \citep{Sankrit2014, Raymond2020a} and $\sim$150~km~s$^{-1}$ in IC~443 \citep{Reach2024}. 
By contrast, slow bow shocks are commonly formed around Herbig-Haro (HH) objects, where jets interact with the dense surrounding ISM \citep{Bally2002, Raga2016}, and these slower shocks, with $v_{\rm s} < 150$~km~s$^{-1}$, are described by the model grid of \citet{Dopita2017}. 
Here, we extend this work by developing model grids for the fast and intermediate-velocity shocks observed in outflows and SNRs, with $v_{\rm s}$ ranging from 100 to 1000~km~s$^{-1}$ in steps of 25~km~s$^{-1}$.

\subsubsection{Abundance}

The choice of the chemical abundance affects the cooling rate and the emission-line spectra from postshock and preshock.
We adopt the ``Galactic Concordance'' for the new shock grids, which coordinates the abundance in stellar atmospheres and nebulae \citep{Nicholls2017}.
This new chemical abundances adopt the measured B star abundance as the reference, and build the abundance of individual element by empirically fitting the stellar atmosphere data. 
Different from the uniform abundance scaling, the ``Galactic Concordance'' abundances apply a non-uniform scaling by fitting the stellar atmosphere data.
These new abundance scaling considers the different enrichment rates of individual elements, which matches with the stellar evolution track better than the uniform scaling models.
We adopt the solar abundance of $\zsun$ = 8.93 \citep{Anders1989}, and the abundance is listed in Table~\ref{tab:abundance}.
The super-solar and sub-solar abundances are obtained by applying this non-uniform scaling described in \cite{Nicholls2017}. 
The metallicities of 0.4$\times\rm Z_\odot$ and 0.2$\times\rm Z_\odot$ are close to those of the ISM in the LMC and SMC, respectively.

\begin{deluxetable}{ccccc}
 \tablecaption{ The ``Galactic Concordance'' Solar Abundance \label{tab:abundance}}
 \tablehead{
 \colhead{Element} & \colhead{12+log(X/H)} & \colhead{} & \colhead{Element} & \colhead{12+log(X/H)}
 }
\startdata
H  & 12.00 && Al &  6.65 \\
He & 11.02 && Si &  7.74 \\
C  &  8.74 && S  &  7.34 \\
N  &  8.11 && Cl &  5.42 \\
O  &  8.93 && Ar &  6.57 \\
Ne &  8.26 && Ca &  6.54 \\
Na &  6.48 && Fe &  7.86 \\
Mg &  7.76 && Ni &  6.54 \\
  \enddata
\end{deluxetable}

\subsection{The \MIII\ Comparison Shock Model Library}

The \MIII\ grid includes the fast shocks with velocities in the range of $v_{shock} = 100-1000~\rm km~s^{-1}$ in steps of 25~km~s$^{-1}$.
The magnetic fields $B$ are from $10^{-4}$ to $100~ \mu G$, and density $n_H$ of 0.01, 0.1, 1.0, 10, 100 and 1000~$\rm cm^{-3}$.
The gaseous chemical abundance is the Magellanic Clouds abundance given by \citep{Russell1992}, the depleted Solar and twice Solar abundance sets based on \citep{Anders1989} used in \citep{Dopita1996}. 
In the \MIII\ comparison model library, we adopt the same parameters but generate the spectra with the \MV\ code.

\section{Shock Structures} \label{sec:postshock}

\begin{figure*}
    \centering
    \includegraphics[width=1\linewidth]{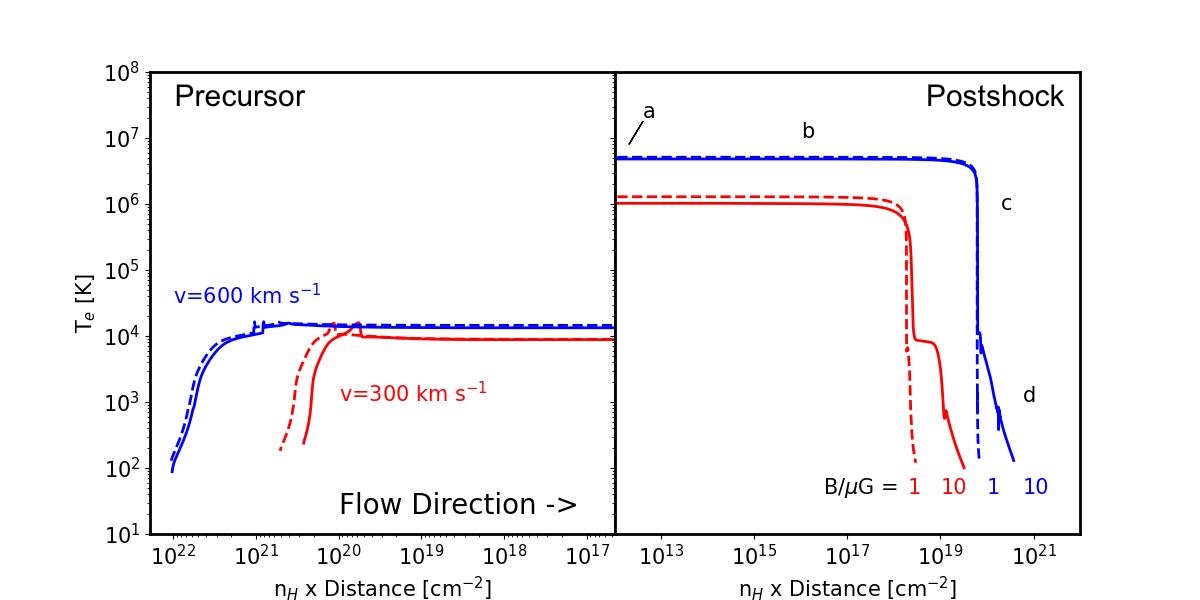}
    \caption{The temperature structure of the entire flow across the precursor and postshock region. The red and blue lines are respectively the 300~$\kms$ and 600~$\kms$ shock models. The dashed lines are the models with weak magnetic strength of 1~$\mu$G and the solid lines are the models with strong magnetic strength of 10~$\mu$G.}
    \label{fig:ShockStruct}
\end{figure*}

Figure~\ref{fig:ShockStruct} presents the temperature structures of $v_{shock}= $300~$\rm km~s^{-1}$ and 600~$\rm km~s^{-1}$ shocks, with density of 1.0~$\rm cm^{-3}$, magnetic strength of 1.0~$\mu$G and 10.0~$\mu$G.
We define four zones in postshock regions: a) the tip of the shock front where the postshock temperature reaches to its maximum; b) the relaxing zone where the gas is relaxed towards the collisional ionization equilibrium after crossing the shock front; c) the cooling zone where the cooling of metal elements happens; d) the recombination tail in which the ionizing photons radiated from the cooling zone dominate the radiation.
The recombination tail is thicker in the higher magnetic field shocks. 
This is because the relative strong magnetic strength suppresses the compression factor of flows and reducing the cooling rate and the production of ionizing photons.

\section{Precursor} \label{sec:precursor}

The temperature, density and ionization state of the precursor gas determine the structure of post-shock. 
In turn, the upstream ionizing photons created from the post-shock cooling zone ionize and heat the preshocked gas, altering the temperature and ionization state of the precursor. 
A self-consistent shock calculation should compute the precursor and post-shock region simultaneously.
In the past, \cite{Shull1979} treat the precursor as an \HII\ region by neglecting the variance of spectrum hardness during radiative transfer, while \cite{Binette1985} approximate the precursor as a single slab photoionized by a mean up-stream field.

\MIII\ grid computes the precursor and shock structures iteratively by following the strategy of \cite{Binette1985} and \cite{Dopita1996}.
A shock model is firstly calculated by an estimate of the ionization state of the precursor.
Then the physical condition of the precursor is updated by computing the photoionization equilibrium gas with the upstream ionization field generated by the shock.
This routine is repeated around four times until both the precursor solution is stable.

\begin{figure*}
    \centering
    \includegraphics[width=1\linewidth]{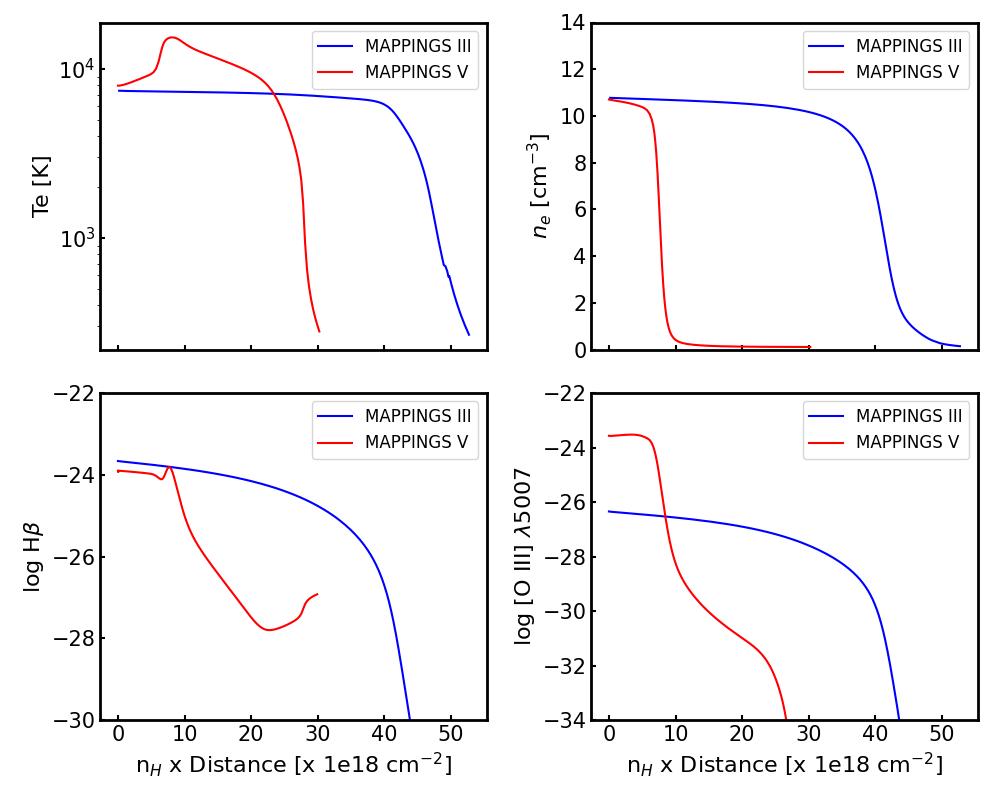}
    \caption{Comparison of precursor structures from the \MV\ (red) and \MIII\ (blue) model with $v_{shock}=$150~$\kms$, $n_H=$10~$\cmc$ and $B=$10~$\mu G$. The \MIII\ precursor model is the photoionization equilibrium precursor ionized by the upstream ionizing photons of the $v_{shock}=$150~$\kms$ shock. The \MV\ precursor model is calculated from a ``multi-zone'' approximation where the ionization and recombination happen within a finite duration between the moments that gas flows into the precursor and the shock front.}
    \label{fig:PrecStruct}
\end{figure*}

However, the precursor does not reach the ionization balance when the ionization front is slower than the shock velocity.
In this case, the ionizing photons are trapped by the incoming neutral preshocked gas.
The ionization state of precursor is a consequence of the ionization and recombination happening within the finite duration between the gas flowing into and out of precursor. 
The temperature is, therefore, hotter in this time-dependent precursor structure as the gas does not effectively cool down within the finite time.
Figure~\ref{fig:PrecStruct} shows the difference between the time-dependent and photoionization equilibrium precursors of the $v_{shock}=150~\kms$ flows.
The time-dependent precursor is shorter than the photoionization equilibrium one as cooling is ineffective at the leading edge of the precursor, far from the shock front.
Table~\ref{tab:precurosrline} gives the precursor emission-line fluxes calculated by different methods.

\begin{table*}
\centering
\caption{Precursor line fluxes calculated through the simple photoionization approximation and multi-zone approach.}
\label{tab:spec_lines}
\begin{tabular}{cccc}
\hline
Wavelength (\AA) & Line Name & Simple Photoionization & Multi-Zone \\
\hline
H$\beta$=100 \\
3726 & \OII & 80.26 & 313.29 \\
3729 & \OII & 115.32 & 450.87 \\
5007 & \OIII & 64.99 & 179.27 \\
6563 & \Ha & 295.56 & 348.06 \\
6583 & \NII & 95.66 & 202.61 \\
6716 & \SII & 56.77 & 131.87 \\
6731 & \SII & 39.35 & 92.13 \\
\hline 
& Flux (${\rm H\beta}$) erg s$^{-1}$ cm$^{-2}$ & 6.619$\times$10$^{-5}$ & 1.290$\times$10$^{-5}$ \\
\hline 
\end{tabular}\label{tab:precurosrline}
\end{table*}

\begin{figure*}
        \centering
    \includegraphics[width=1\linewidth]{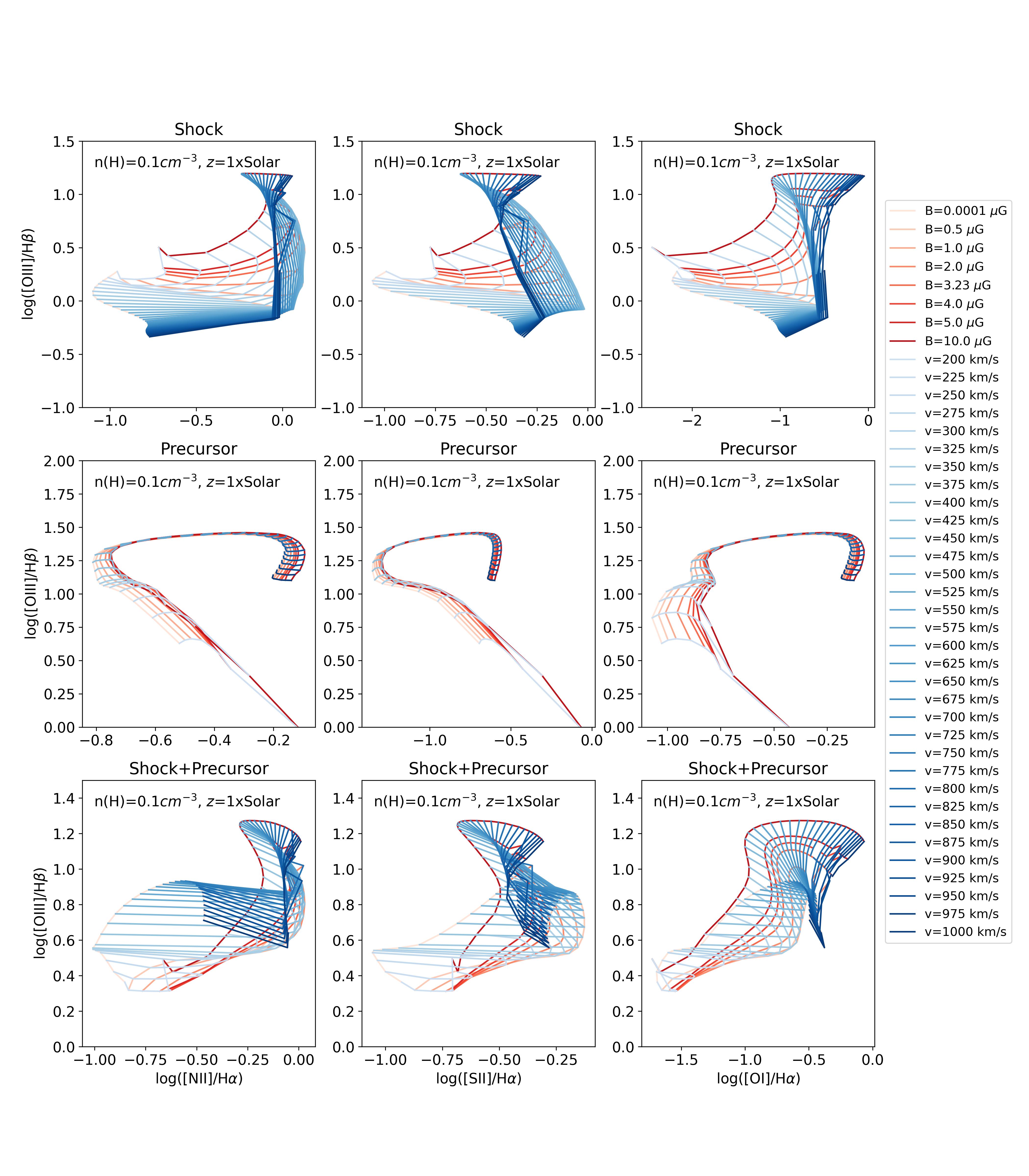}
    \caption{The shock grids on the BPT diagram for the models with Solar metallicity, $n_H$=0.1$cm^{-3}$. {\bf Top:} The grids of the shock spectra. The red lines represent varying magnetic field strengths and the blue lines represent varying shock velocities. {\bf Middle:} The grids of the precursor spectra. {\bf Bottom:} The grids of the shock and precursor spectra.}
    \label{fig:grid_n0.1}
\end{figure*}

\begin{figure*}
        \centering
    \includegraphics[width=1\linewidth]{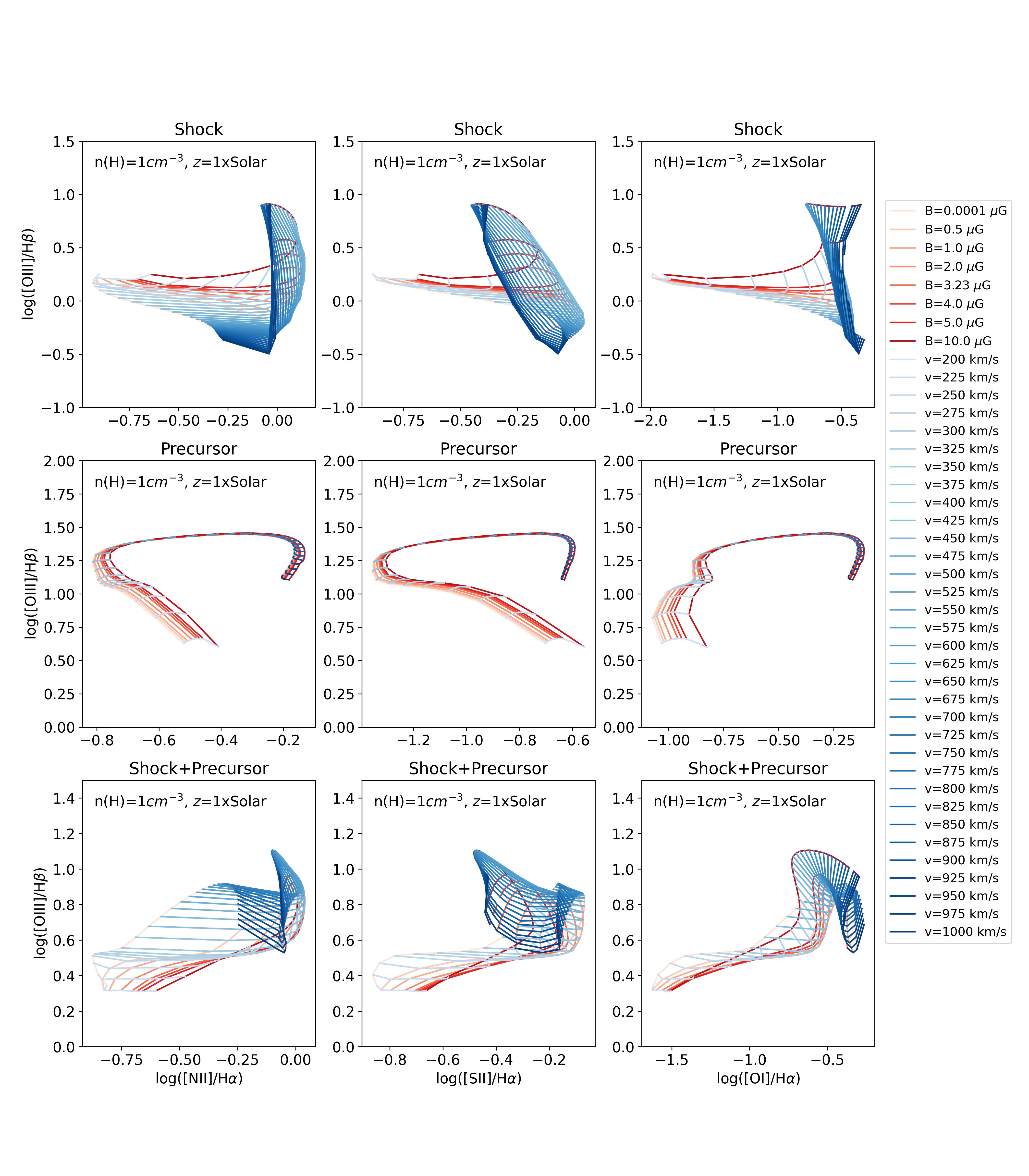}
    \caption{Same to Figure~\ref{fig:grid_n0.1} but for the $n_H$=1$cm^{-3}$ models.}
    \label{fig:grid_n1}
\end{figure*}

\begin{figure*}
        \centering
    \includegraphics[width=1\linewidth]{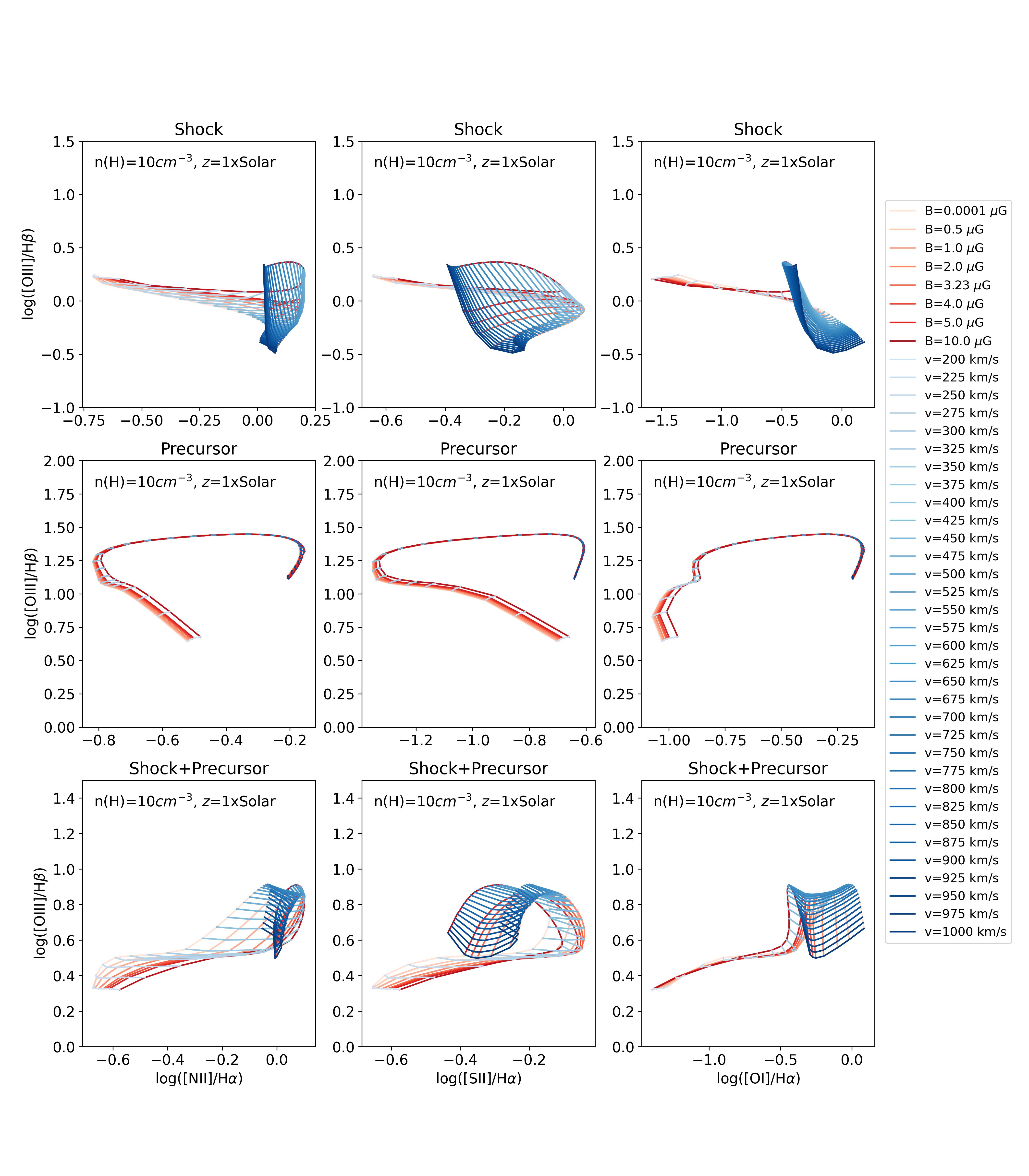}
    \caption{Same to Figure~\ref{fig:grid_n0.1} but for the $n_H$=10$cm^{-3}$ models}
    \label{fig:grid_n10}
\end{figure*}

\begin{figure*}
        \centering
    \includegraphics[width=1\linewidth]{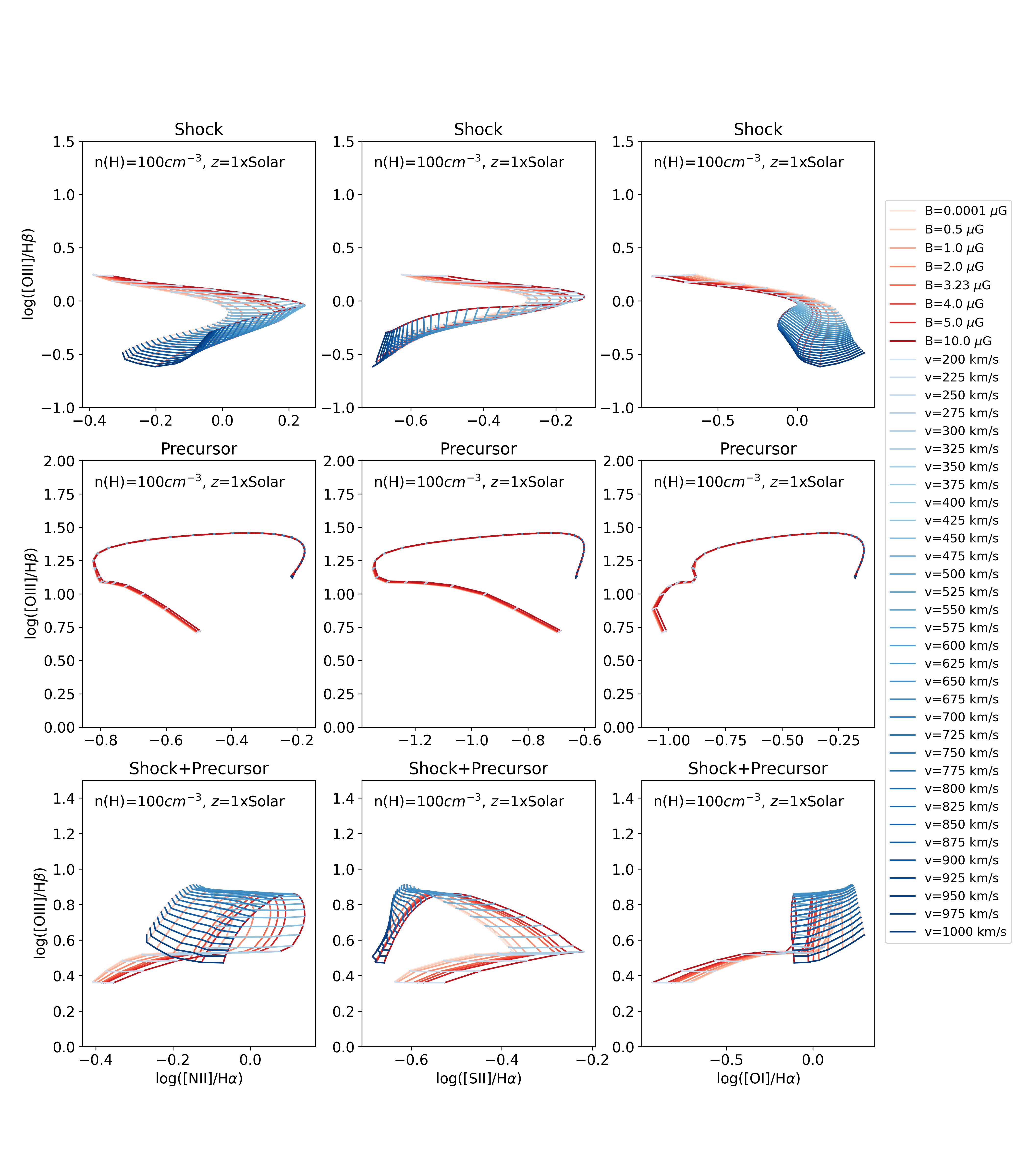}
    \caption{Same to Figure~\ref{fig:grid_n0.1}, but for the $n_H$=100$cm^{-3}$ models}
    \label{fig:grid_n100}
\end{figure*}

\section{Shock Emission Lines} \label{sec:line}

The synchronous time-dependent solution of shock flows largely improves the accuracy of the prediction of emission-line spectra in both precursor and shocked regions.
The changes in the ionization state of the preshock gas arriving at the shock front alter the physical states of post-shock regions and the emission-line created in cooling and recombination zones.
The high ionization level of gas suppresses the cooling efficiency of post-shock regions, reducing the cooling efficiency and sustaining the elevated temperature until the post-shocked gas reaches the ionization balance and the collisional cooling becomes efficient.
Then, the cooling spectra going upstream and downstream create the photoionized region in precursors and recombination zones where the emission-lines are produced.

\subsection{The optical emission-line grids}

The optical emission-lines are the most useful probes of the shocks from the supernova remnants and the AGN outflows.
The collisional-excited lines, \NII, \SII, \OI, are the featuring optical lines of shocks which are elevated as the fraction of shock-to-\HII\ region emissions increases.
These lines are used in conjunction with \OIII\ and Balmer lines to form the diagnostic diagrams to separate shocks from other ionization mechanisms, like the Baldwin-Phillips-Terlevich diagram \citep[the BPT diagram,][]{Baldwin1981,Veilleux2005} and the WHAN diagram \citep{Fernandes2011}.
Figure~\ref{fig:grid_n0.1} to \ref{fig:grid_n100} present the locus of the shock emission-line ratios on the BPT diagram by varying the magnetic field from 0.0001~$\mu G$ to 10~$\mu G$ and the shock velocity from 200~$\kms$ to 1000~$\kms$.
The preshocked gas density and the chemical abundance of the models are fixed in each figure.

In the shocks with a preshock density below 10 $\cmc$, the \OIIIHb\ ratio is not sensitive to the shock velocity, but rather to the magnetic strength.
This is because for these fast shocks, the oxygen is fully ionized twice before the post-shocked gas cools down.
When the magnetic strength is stronger, part of the shock energy is stored in the Alfv\'{e}n wave and released in the enlarged recombination zone.
As a consequence, the \OIIIHb\ ratio increases with the stronger magnetic strength.
When the gas density increases, ram pressure, $P_{ram}=\rho v^2$, becomes dominant over magnetic pressure, $P_{B}=B^2/8\pi$.
Therefore, the spread of the \OIIIHb\ ratio is smaller in high-density, low-velocity shocks than in the low-density, high-velocity shocks.
In the shocks with a preshock density of 100 $\cmc$, the line-ratios are driven by the shock velocity. 
The magnetic field has negligible effect on shock emission-line ratios.

In Figures~\ref{fig:grid_n0.1} through \ref{fig:grid_n100}, the shock model grids fold back in the \NIIHa, \SIIHa, and \OIHa ratios at high shock velocities.
This is because the density-weighted ratio of the cooling zone length to the size of the entire post-shock increases as the shock velocity increases from 200~$\kms$ to several hundred $\kms$.
As the shock velocity increases from several hundred $\kms$ to 1000~$\kms$, the post-shock region grows in size more rapidly than the cooling zone length, leading to a decrease in the \NIIHa\, \SIIHa\, and \OIHa\ ratios.

The line-ratios of precursors are determined by shock velocities.
As defined in \cite{Sutherland2017}, the precursor emission is determined by the shock-precursor parameter, $\Psi=Q/v_{s}$, where $Q$ is the ionization parameter of the upstream ionizing photons arriving at the shock front and $v_{s}$ is the shock velocity.
The shock-precursor parameter describes the speed of ionization front relative to the speed of shock flows, and $\Psi\propto v_{s}^\alpha$, where the $\alpha\gg2$ at $v_{s}\sim50~\kms$ and approaches to 2 at high shock velocities.
A detailed discussion of the dependence of the line-ratios on the shock parameters can be found in \cite{Sutherland2017}.

\section{Comparison between MAPPINGS III and MAPPINGS V models} \label{sec:comp_MIII}

The improvement from \MIII\ to \MV\ are twofold: the computation of precursor and the adoption of atomic data. 
\MIII\ approximates the precursor as a single slab of fully photoionized gas just before the gas passes the shock front.
This approximation works when the upstream ionizing fronts travel faster than the shock fronts to allow for the gas reaching the photoionization equilibrium.
When the shock velocity is greater than the velocity of ionizing front, the gas will travel through the precursor and enter the shock front before reaching the ionization-recombination balance.
\MV\ abandons the equilibrium assumption by calculating the ionization and recombination processes between the moment when the gas enters the precursor ionizing front and when the gas enters the shock front.
This avoids to over-ionizing the precursor gas when the shock velocity is slower than the speed of upstream ionizing front, as shown in Table~\ref{tab:comp_ionization}.

\begin{figure*}
    \centering
    \includegraphics[width=1\linewidth]{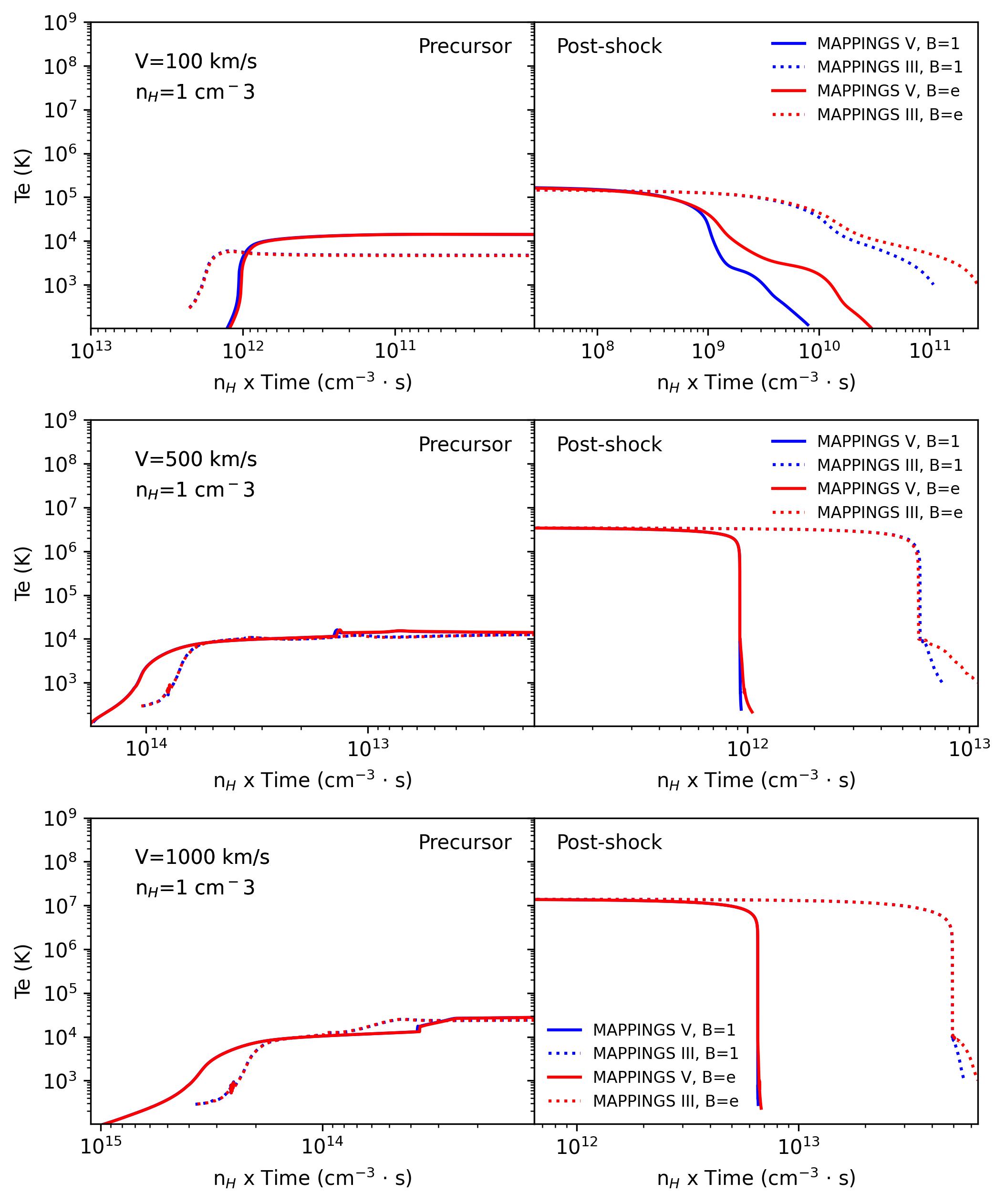}
    \caption{Comparison of the temperature structure between \MV\ and \MIII\ models with magnetic strength $B=1~\mu G$ and $B=3.23 \mu G$ (B=e).}
    \label{fig:comp_m3_m5}
\end{figure*}

\MV\ uses the CHIANTI~10 database while \MIII\ uses a collective of atomic data measured before the 1990s \citep{Sutherland1993}.
The changes of atomic data alter the cooling efficiency.
In conjunction with the changes of ionization states in preshocked gas, the \MV\ post-shock regions are shorter than the \MIII\ regions because the \MV\ models cool down more efficiently than the \MIII\ models, as it is shown in Figure~\ref{fig:comp_m3_m5}.

\begin{figure*}
    \centering
    \includegraphics[width=1\linewidth]{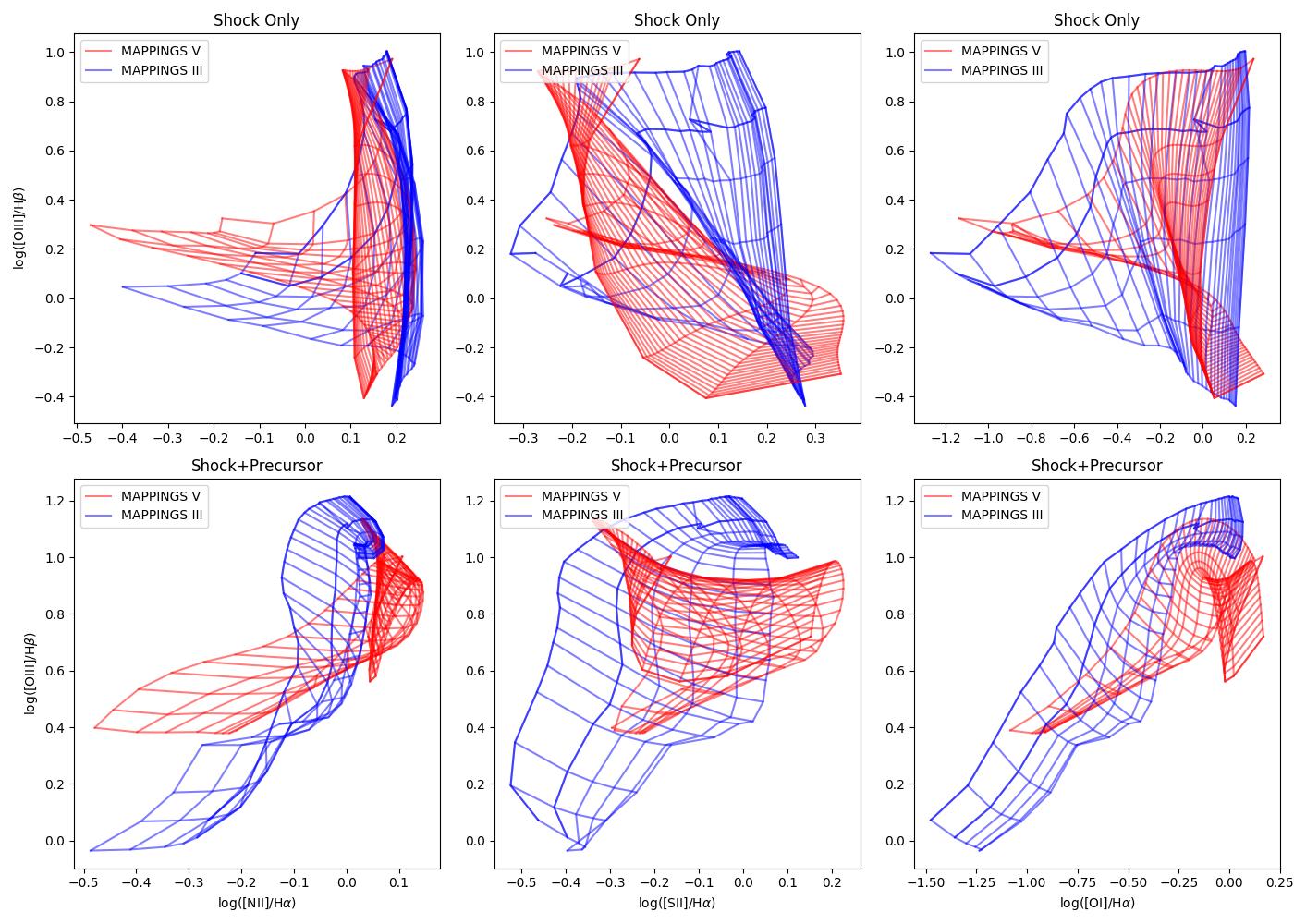}
    \caption{Comparison between \MV\ and \cite{Allen2008} shock grids.}
    \label{fig:allen_compare}
\end{figure*}

\begin{figure*}
    \centering
    \includegraphics[width=1\linewidth]{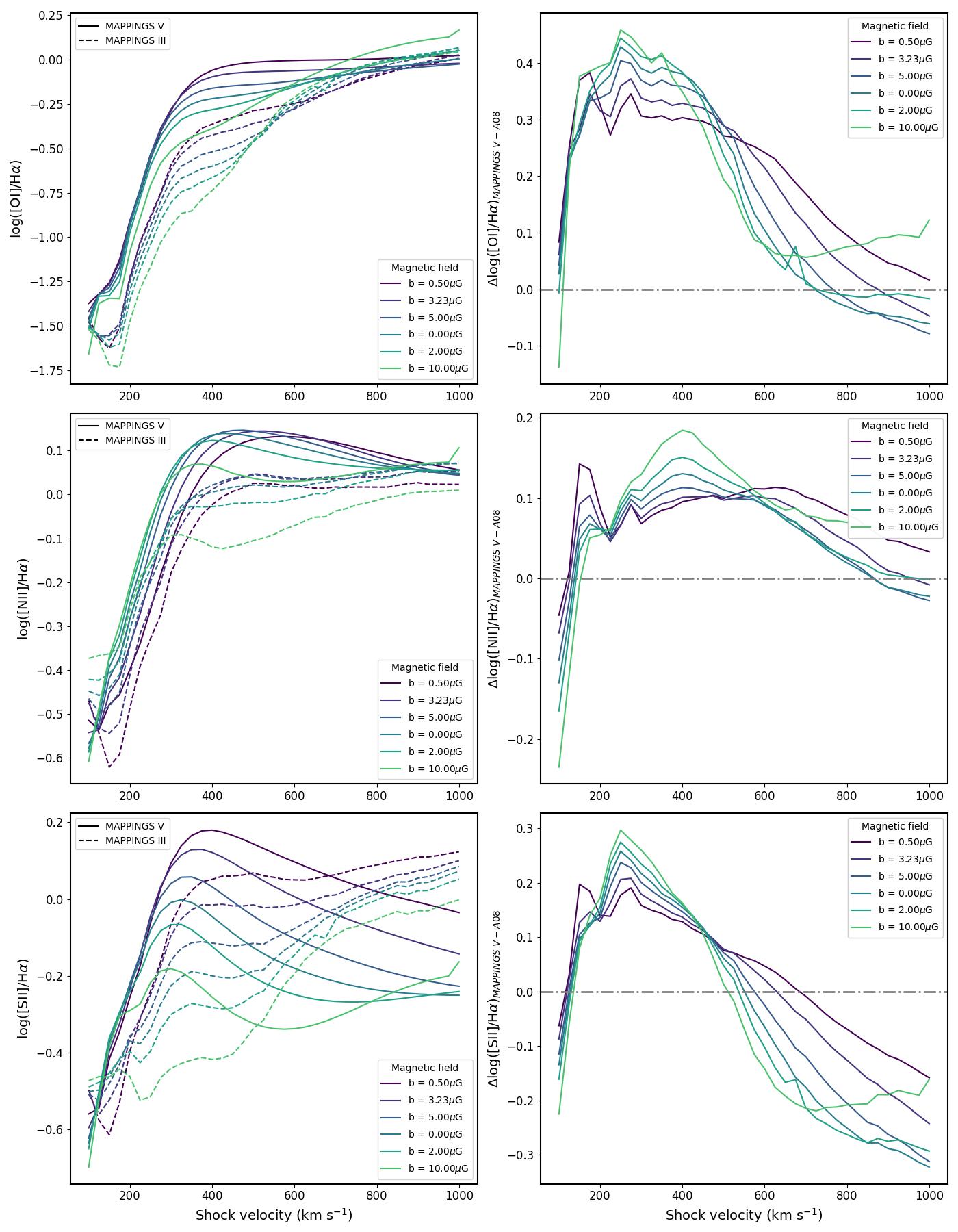}
    \caption{The difference between the \MV\ and \MIII\ models in shock+precursor emission-line ratios as a function of shock velocity. {\bf Left:} The shock+precursor emission-line ratios change with shock velocity in the \MV\ model (solid line) and \MIII\ model (dashed line). {\bf Right:} The difference between the \MV\ and \MIII\ models in shock+precursor emission-line ratios. The dotted-dashed line indicates there is no difference between the two models.}
    \label{fig:spl_change}
\end{figure*}

Figure~\ref{fig:allen_compare} presents the \MIII\ and \MV\ shock spectral grids, which have the same chemical abundance.
The two grids behave similar trend on the \NIIHa-BPT diagram but the \OIIIHb, \NIIHa\ and \OIHa\ ratios span a wider range in the \MIII\ precursor+shock models than the \MV\ models.
The \SIIHa\ and \OIHa\ ratios have different dependence on the shock velocity between the \MIII\ and the \MV\ models.
As shown in Figure~\ref{fig:spl_change}, the total \OIHa\ ratio increases monotonically with shock velocity in \MIII\ models, while in \MV\ models, the \OIHa\ ratio remains flat with shock velocity at $v_{shock}>400~\kms$.
The total \OIHa\ ratios of the shock and precursor in the \MV\ models are larger than those of the \MIII\ models by up to 0.4~dex.
In \MIII\ models, the \SIIHa\ ratio slightly increases with shock velocity when $v_{shck}>400~\kms$, while in \MV\ models, the \SIIHa\ decreases with increasing shock velocity when $v_{shock}>400~\kms$.
Therefore, the difference of the total \SIIHa\ ratios between the \MV\ and \MIII\ models is around 0.2 -- 0.3~dex at $v_{shock}\sim300~\kms$, and decreases to -0.3 -- -0.2~dex at $v_{shck}\sim1000~\kms$.

\begin{table*}[htbp]
\centering
\caption{The ionization fraction of H and He before the gas crosses shock fronts for MAPPINGS III and V Models}
\label{tab:comp_ionization}
\setlength{\tabcolsep}{3pt}
\begin{tabular}{lcccccc}
\toprule
 & \multicolumn{6}{c}{Model} \\
\cmidrule(lr){2-7}
n(H) & \multicolumn{6}{c}{1 $\cmc$} \\
$v_{\text{shock}}$& \multicolumn{2}{c}{100 km s$^{-1}$} & \multicolumn{2}{c}{500 km s$^{-1}$} & \multicolumn{2}{c}{1000 km s$^{-1}$} \\ 
B & \makecell{1 $\mu$G} & \makecell{3.23 $\mu$G} & \makecell{1 $\mu$G} & \makecell{3.23 $\mu$G} & \makecell{1 $\mu$G} & \makecell{3.23 $\mu$G} \\
\midrule
\multicolumn{7}{c}{\textbf{MAPPINGS III}} \\
H\,I & 0.039 & 0.039 & 0.000 & 0.000 & 0.000 & 0.000 \\
H\,II & 0.961 & 0.961 & 1.000 & 1.000 & 1.000 & 1.000 \\
He\,I & 0.163 & 0.165 & 0.000 & 0.000 & 0.000 & 0.000 \\
He\,II & 0.832 & 0.830 & 0.021 & 0.020 & 0.005 & 0.005 \\
He\,III & 0.005 & 0.005 & 0.979 & 0.980 & 0.995 & 0.995 \\
\midrule
\multicolumn{7}{c}{\textbf{MAPPINGS V}} \\
H\,I & 0.630 & 0.676 & 0.001 & 0.001 & 0.000 & 0.000 \\
H\,II & 0.370 & 0.324 & 0.999 & 0.999 & 1.000 & 1.000 \\
He\,I & 0.686 & 0.719 & 0.000 & 0.000 & 0.000 & 0.000 \\
He\,II & 0.313 & 0.280 & 0.028 & 0.027 & 0.005 & 0.005 \\
He\,III & 0.001 & 0.001 & 0.972 & 0.973 & 0.995 & 0.995 \\
\bottomrule
\end{tabular}
\end{table*}

\section{Observational Comparisons}

\subsection{Galaxy spectra}\label{sec:comp_sdss}

\begin{figure*}
        \centering
    \includegraphics[width=1\linewidth]{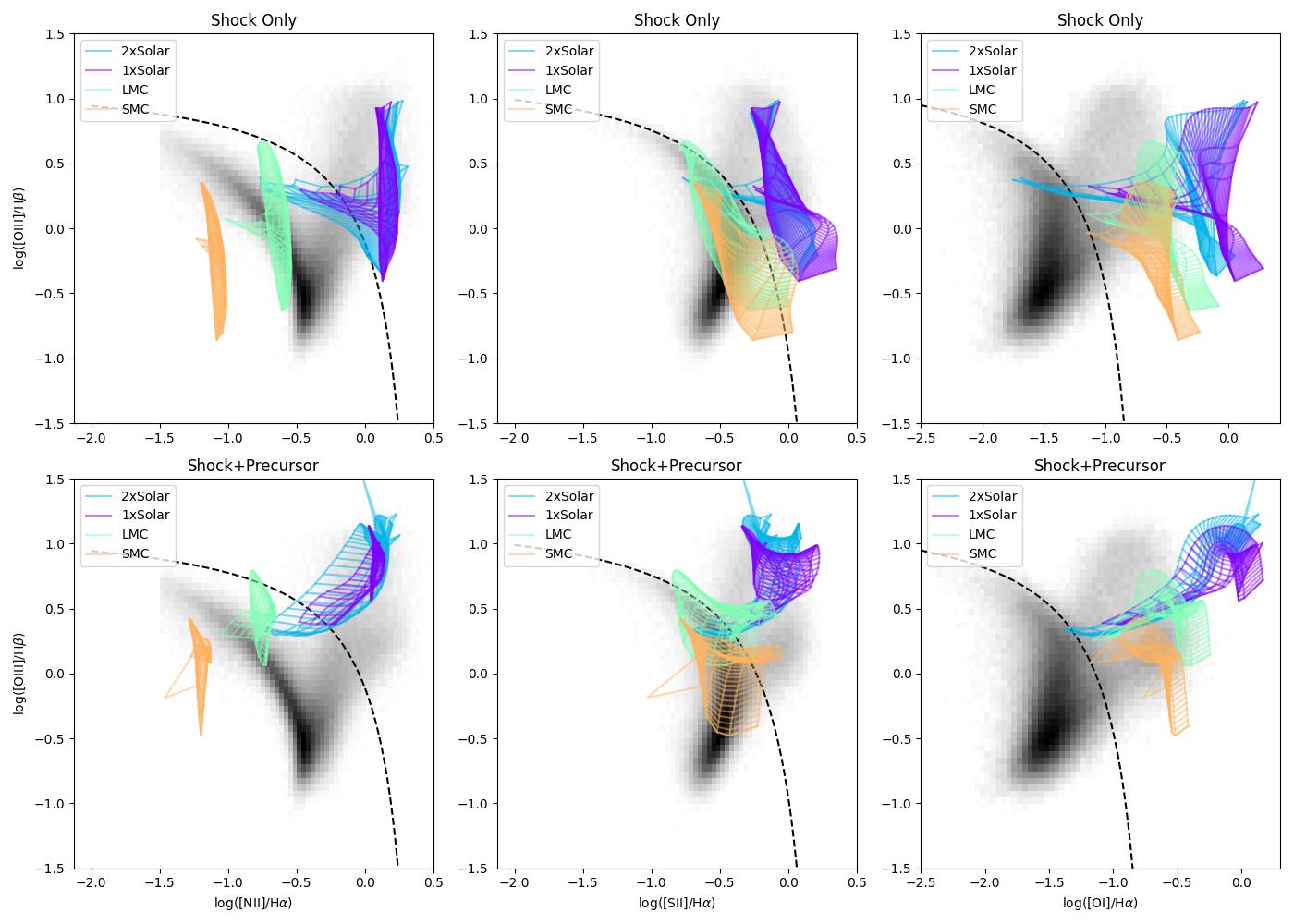}
    \caption{Comparison between the \MV\ models and SDSS DR7 galaxy spectra. The background gray points are the SDSS DR7 emission-line galaxies with emission-line signal-to-noise ratio greater than 3. The different color grids are the \MV\ models with different chemical abundances. The dashed line on each panel is the maximal line-ratios given by \HII\ region models listed in \cite{Kewley2006}.}
    \label{fig:sdss_compare}
\end{figure*}

Emission-line galaxies are classified as star-forming galaxies, AGN and low-ionization regions (LINERs).
Shock is an ionization mechanism candidate to produce the low-ionization regions.
Figure~\ref{fig:sdss_compare} shows shock models compared against emission-line ratios of the SDSS galaxies \citep[Sloan Digital Sky Survey,][]{Abazajian2009} on the BPT diagrams.
The SDSS DR7 is a volume-limited sample of galaxies spanning redshift of $0<z<0.4$. 
The galaxies are clustered into two branches on the BPT diagram.
The left branch is the ``star-forming'' branch where galaxies are dominated by the ionization of star formation \citep{Kauffmann2003,Kewley2006}, while the right branch is powered by non-star-forming processes, like AGNs, shocks and hot evolved stars \citep{Byler2019}.
We compare the data with four shock grids of 2$\times Zsun$, 1$\times Zsun$, LMC and SMC chemical abundances.

At the scale of galaxies, shocks are majorly from the AGN outflows.
The recent spatially-resolved study of the local AGNs finds that the fast shock models agree to the outflow-only emission after removing the contribution of \HII\ regions \citep{DAgostino2018}.
For integrated galactic spectra, the emission is a hybrid of the contribution from shock and \HII\ regions. 
The shock models can roughly demonstrate where shocks dominate the emissions.
As shown in Figure~\ref{fig:sdss_compare}, the precursor+shock and shock-only grids agree to the AGN branch on the BPT diagram.
However, on the \NIIHa-BPT diagram, the LMC- and SMC-like low-metallicity shocks are hardly separated from the \HII\ regions as they merge with the star-forming branch.
However, on the \OIHa-BPT diagram, it is clear that the shock models agree to the location of the AGN branch, which is a good indicator for the AGN outflows studies.

\subsection{Supernova remnant in external galaxies}\label{sec:comp_snr}

\begin{figure*}
        \centering
    \includegraphics[width=1\linewidth]{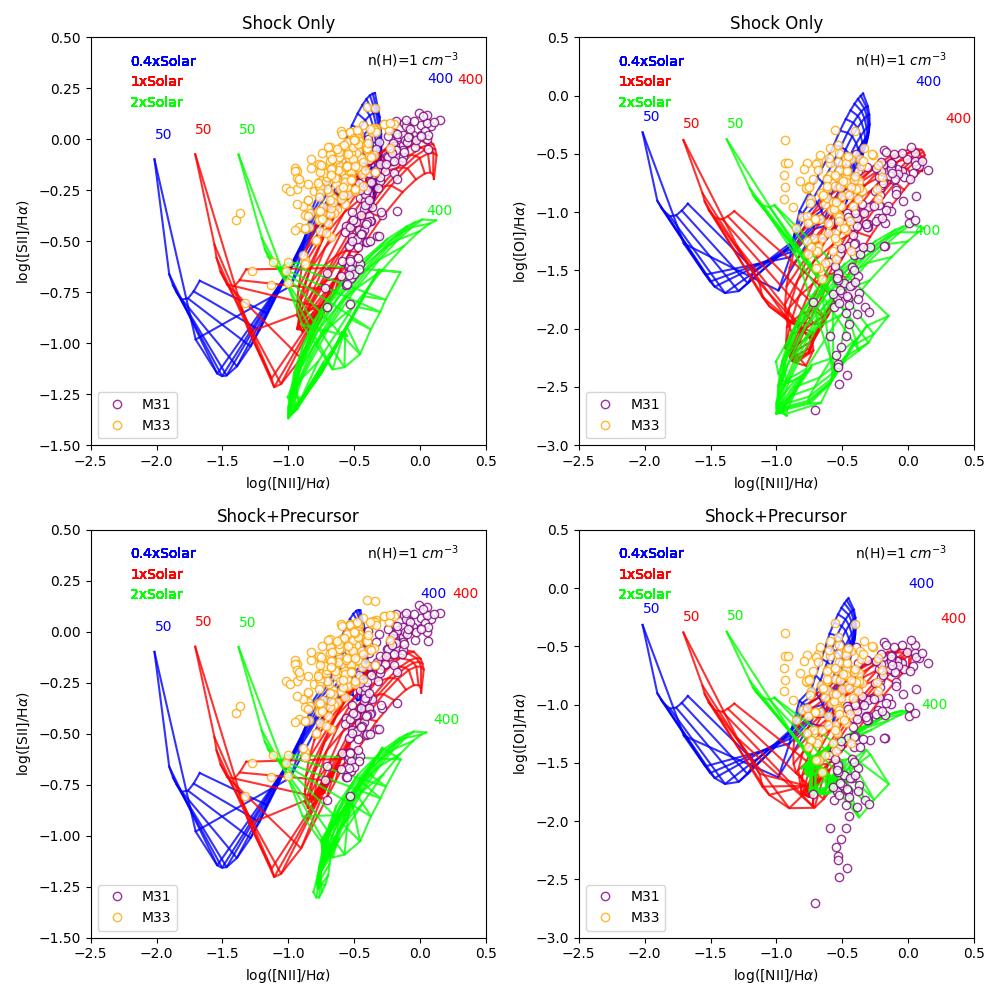}
    \caption{Comparison between the \MV\ models and the supernova remnant sample in M31 and M33 on the \NIIHa-\SIIHa\ diagram (left panels) and \NIIHa-\OIHa\ diagram (right panels). {Top:} The data against the shock-only emission-line ratios. The blue, red and green grids correspond to the sub-solar (0.4$\times \rm Z_\odot$), solar and super-solar (2$\times \rm Z_\odot$) abundance. We only display the $50~< v_{shock}<~400~\kms$ models with density of $n_H=1~\cmc$. The slow and fast velocity ends are indicated by the values of velocity. The purple circles are the SNR candidates in M31 and the orange circles are those in M33. {\bf Bottom:} The same emission-line diagram as the top panels but the models are the shock+precursor emission-line ratios.}
    \label{fig:snr_compare}
\end{figure*}

Figure~\ref{fig:snr_compare} compares the SNR data to the model grids on the \NIIHa-\SIIHa\ and \NIIHa-\OIHa\ diagram. 
To fit the shock velocity in SNRs, the grids include 50~$\kms$ to 400~$\kms$ shock models.
The data are the MMT spectroscopy of the SNR candidates in M31 \citep{Caldwell2025} and in M33 \citep{Long2018}, which are selected on the basis of \SIIHa\ ratio, and largely remove the contamination from \HII\ regions.
We select the $\rm n_H=1~\cmc$ grids as it is the typical ISM density \citep{Spitzer1978}.
We only include the 0.4$\times \rm Z_\odot$, 1.0$\times \rm Z_\odot$ and 2.0$\times \rm Z_\odot$ metallicity, as the metallicity is around 1.3$\times \rm Z_\odot$ in M31 \citep{Zurita2012} and 0.5$\times \rm Z_\odot$ in M33 \citep{Rosolowsky2008}. 
The M31 and M33 SNR data agree with the grids of velocity $v_{shock}\gtrsim85~\kms$, and the difference of the line-ratios between the shock-only and the precursor+shock models are not pronounced.
The contribution of the precursor to the total SNR spectrum is unclear yet as the precursor may not be fully developed in real SNRs.
The \NIIHa-\SIIHa\ and \NIIHa-\OIHa\ diagrams are good diagnosis for separating the SNRs of different metallicities.
Agreeing to the measured metallicity, the best-fit models for the M31 and M33 SNRs are the 1$\times \rm Z_\odot$ and the 0.4$\times \rm Z_\odot$ grids, respectively.
Recent high-velocity-resolution spectroscopic study shows that the interaction between SNR and its N-enriched circumstellar medium may change the line fluxes of shock waves \citep{Caldwell2025}.
Further studies of SNR emission-lines are needed, but the grid itself can give a reliable reference to catch the general properties of SNRs.

\section{Conclusion}\label{sec:conclusion}

In this paper, we present a new $\MV$ shock-precursor spectral library of complete shocks. The library covers a large parameter range of the shock velocity, density, magnetic strength, and chemical abundance and is intended to be useful for studying shocked gas in supernova remnants and AGN outflows. We present the shock-precursor structures and the integrated emission-lines of shocks and precursors. This library is to be preferred to that  \cite{Allen2008} as a result of the improvements made to Mappings since that grid was created, most notably the incorporation of a fully-time-dependent ionization and self-consistent shock-precursor calculation.

On the BPT diagram, the \MV\ shock grids are more compact than the  \MIII\ shock grids. The precursor emissions are overestimated in \MIII\ grids. The simple photoionization equilibrium assumption hardly solve the structures and emissions of precursors.

We further compare the model emission-line ratios with the observations and have the following findings.
\begin{itemize}
    \item We compare the model emission-line ratios with the SDSS galaxy sample, showing that the shock models agree to the AGNs branch on the BPT diagram. 
    This implies that the shocks in AGN outflows are a candidate of the dominating ionizing source.
    We further suggest that the \OIHa-BPT diagram is the best diagnostic for separating the shock-dominating sources from the star-forming galaxies because the shock model grids clearly separate from the star-forming branch on the BPT diagram.
    \item We compare the shock grids with the SNRs in M31 and M33.
    We find that the shock velocity in these SNRs is faster than around 85~$\kms$, and the metallicity of the SNRs is consistent with their host galaxies.
\end{itemize}

While quantitatively different from results obtained with \MIII, the overall trends are similar to that derived using the \cite{Allen2008} grid.

The current \MV\ shock library adopts the plane-parallel geometry in shocked flows. 
A shock models with 3D geometries is the future to compare against the realistic shocks in the data. 

{\bf The shock and precursor spectral library} is available in Zenodo, at
\dataset[doi:10.5281/zenodo.20716483]{https://doi.org/10.5281/zenodo.20716483}.

 {\bf The \MV\ code} is regularly updated at \url{https://github.com/MAPPINGS-Code/MAPPINGS-V}, and the version used for the calculations in this paper is available at \dataset[doi:10.5281/zenodo.24097629]{https://doi.org/10.5281/zenodo.24097629}.

\begin{acknowledgments}

Y.F.J. acknowledges the computational resources at the Westlake High-Performance Computing Center.
Y.F.J. and Y.S. acknowledge the support from the Startup funding of New-joined PI of Westlake University.
This work acknowledges the support from the National Natural Science Foundation of China (NSFC grants 12121003, 12141301, 12333002) and the National Key R\&D Program of China No. 2023YFA1608204.

\end{acknowledgments}

\appendix
\section{\MV\ shock-precursor library}

We construct two structured and queryable SQLite databases of the shock
and precursor models, \texttt{MAPPINGS\_V\_521\_shock.db} and
\texttt{MAPPINGS\_V\_521\_precursor.db}.
Each database stores the model grid in two linked tables:
\begin{enumerate}
  \item A ``Model'' table holding one row per model with its
        \texttt{model\_id} and its physical parameters (preshock
        hydrogen density $n_H$ in $\cmc$, shock velocity in $\kms$,
        magnetic field $B$ in $\mu G$, metallicity $Z/Z_\odot$, the
        corresponding chemical abundance file, and the integrated
        \Hb\ flux in \si{erg.s^{-1}.cm^{-2}}).
  \item A ``Spectra'' table holding the associated emission-line
        spectra (wavelength in \AA, flux (\Hb\ = 1.00), species, and
        line kind) linked to models via their \texttt{model\_id}.
\end{enumerate}
Only lines with a flux relative to \Hb\ greater than $10^{-3}$ are
retained.



\bibliography{sample701}{}
\bibliographystyle{aasjournal}

\end{CJK*}
\end{document}